\documentstyle[11pt,newpasp,twoside]{article}
\include{psfig}
\def\edcomment#1{\iffalse\marginpar{\raggedright\sl#1\/}\else\relax\fi}
\reversemarginpar

\begin{document}
\title{Adaptive Optics in Star Formation}
 \author{Wolfgang Brandner}
\affil{Max-Planck-Institut f\"ur Astronomie, K\"onigstuhl 17, D-69117 Heidelberg, Germany}

\begin{abstract}
Over the past ten years, the concept of adaptive optics has evolved from
early experimental stages to a standard observing tool now available at 
almost all major optical and near-infrared telescope facilities. Adaptive 
optics will also be essential in exploiting the full potential of the 
large optical/infrared interferometers currently under construction. Both 
observations with high-angular resolution and at high contrast, and with 
a high point source sensitivity are facilitated by adaptive optics.
Among the areas which benefit most from the use of adaptive optics are
studies of the circumstellar environment (envelopes, disks, outflows), 
substellar companions and multiple systems, and dense young stellar 
populations. This contribution highlights some of the recent advances in
star formation studies facilitated by adaptive optics, and gives a brief
tutorial on optimized observing and data reduction strategies.
\end{abstract}

\section{AO - (emerging) main-stream observing technique for SF studies}

More than twenty contributions to IAU Symp.\ 221 (many of which can be found
in the present proceedings) are based at least in part on observations with
ground-based telescopes equipped with Adaptive Optics (AO). Hence it is
fair to say that AO has already been adopted as a main-stream observing 
technique for high-resolution star formation (SF) studies.
The SF research topics addressed by AO observations range from studies
of young substellar companions and the circumstellar environment of
young stars (disks \& outflows), over studies of sub-stellar and 
stellar binaries, to studies of massive star forming sites and environments 
(including Galactic and extra-galactic starbursts).

\section{AO Basics: Observing facilities \& glossary of AO terms}

\subsection{Observing facilities}

AO is the real-time measurement and correction of the effect of atmospheric 
turbulence on an electromagnetic wavefront from a distant (astronomical)
source. The temporal stability of an AO correction facilitates long
integration times, and hence increases the limiting sensitivity of
any detector system with a finite read-noise.
By now, many of the 3m- to 10m-class optical/NIR telescopes are equipped
with AO. Astronomical AO systems are using Shack-Hartmann-,
Curvature- or Pyramid-type wavefront sensors with 19 to $\approx$1000 sensing
elements, and deformable mirrors with a corresponding number of actuators.
Wavefront sensing is in general done in the optical, though some of the newer
systems are capable of sensing low-order tip-tilt aberrations also in the
near-infrared, and at least one system (NACO at the ESO VLT) is equipped with
a full-fledged NIR wavefront sensor. Science instruments in general cover
the wavelength range from 1 to 5\,$\mu$m,  and typical instrument modes
cover direct imaging, polarimetry, long-slit and integral field spectroscopy
(including scanning Fabry-Perots), coronagraphy, and differential 
imaging.

\subsection{Brief Glossary of AO related \& relevant terms}

Since AO systems do a real-time analysis and compensation of atmospheric 
wavefront aberrations, the properties of the correction, and hence the quality 
of the science image, vary as a function of atmospheric (turbulence) parameters.
A basic understanding of the physical parameters characterising the
atmospheric turbulence and the (instrumental) limitation of a given
AO system is a prerequisite for designing successful AO observing runs, and
for devising an optimized data reduction and analysis strategy. \\

Definitions of common AO terms:
\begin{itemize}
\item[$\bullet$]{\bf Strehl ratio (SR)}: the ratio between the peak flux of an
observed point-spread function (PSF) and a perfect (diffraction limited)
PSF normalized to the same flux

\item[$\bullet$]{\bf Fried parameter}: r$_{\rm 0}$ corresponds to the size of the 
atmos.\ turbulence cell projected to the ground layer. It scales with wavelength as r$_{\rm 0}$ 
$\propto$ $\lambda ^{6/5}$

\item[$\bullet$]{\bf Seeing angle}: It is defined as $\beta = \lambda / r_{\rm 0} \propto \lambda ^{-1/5}$
(0.5$''$ seeing at $\lambda = $ 500\,nm corresponds to r$_{\rm 0} = $ 0.2\,m)

\item[$\bullet$]{\bf Coherence time}: $\tau_{\rm 0}$ is the lifetime of 
(atmospheric) speckles. It scales with wavelength as $\tau_{\rm 0}$ 
$\propto$ $\lambda ^{6/5}$

\item[$\bullet$]{\bf Isoplanatic Angle}: $\Theta _{\rm 0}$ is the angular
distance from the reference source used for wavefront sensing, at which the
Strehl ratio drops by 1/e. It scales with wavelength as $\Theta_{\rm 0}$
$\propto$ $\lambda ^{6/5}$ (i.e., an isoplanatic angle size of 3$''$ at a
wavelength of 500\,nm corresponds to $\Theta = 18''$ at $\lambda = $2.2\,$\mu$m

\end{itemize}

It is important to keep in mind that the atmospheric turbulence properties
r$_{\rm 0}$, $\tau _{\rm 0}$, and $\Theta _{\rm 0}$ can vary independently
of each other. In particular, $\beta$ and $\Theta _{\rm 0}$ may originate in 
different layers of the atmosphere.
In general, observations at longer wavelengths benefit from better seeing
($\beta$) and longer coherence time ($\tau _{\rm 0}$), and hence result 
in a better AO correction and higher Strehl ratio.

\subsection{How to design a successful AO observing run}

The four main ingredients for designing and executing a successful AO
observing run are {\bf i)}
Select a {\bf bright, compact reference source}: NACO at the ESO 
VLT, e.g., gives partial correction for reference sources as faint as 
V$\le$17\,mag, but full AO correction requires V$\le$12\,mag. In the limit,
only low-order aberration like tip-tilt are corrected, resulting in a low SR.
{\bf ii)}
{\bf Small angular separation} between the reference source and
the science target (i.e. within the isoplanatic angle) is mandatory to
achieve at least 37\% (1/e) of the on-axis SR.
{\bf iii)} {\bf Good atmospheric conditions} (large r$_{\rm 0}$, $\tau _{\rm 0}$) 
become increasingly important for observations at shorter wavelengths (going
from \dots $\rightarrow$ L $\rightarrow$ K $\rightarrow$ H $\rightarrow$ J $\rightarrow$ Z $\rightarrow$ I $\rightarrow$ \dots). 
Hence it is advisable to choose
service mode observations wherever this mode if offered.
and {\bf iv)}
A {\bf Reference PSF} is important for the interpretation of
AO observations of extended sources, and for high-contrast applications.

\subsection{AO specific effects to watch out for during data reduction}

AO data are subject to angular and temporal variations of the PSF,
as well as differential atmospheric refraction, which at 8\,m- to 10\,m-class
telescopes can be of the same order of magnitude as the diffraction limited PSF size.

\subsubsection{Temporal PSF variations} 

Temporal variations of the PSF imply that the quality of individual science
frames has to be checked before combining or averaging the frames, and that
in particular in the case of marginal data, an image selection has to be 
applied (Tessier et al.\ 1994). In crowded stellar fields, the angular 
variation of the PSF can
be measured and modelled, and then taken into account in the course of
the data analysis (e.g., using a linearly varying PSF in DAOPHOT, or the
new implementation of the StarFinder code). In the case of morphological
studies of extended sources like
deeply embedded Young Stellar Objects, galaxy clusters, and lensing studies,
a knowledge of the atmospheric turbulence profile is required for a proper
interpretation of the data. Current AO systems are not well suited
for ``wide-field'' morphological studies. These studies therefore have
to be deferred until multi-conjugated AO (MCAO) systems become operational.

\subsubsection{Speckle noise and high-contrast AO}

AO systems operate in a ``closed-loop'' configuration. The
wavefront sensor signal corresponds to an error signal, i.e., the difference
between the corrections applied by the deformable mirror and the actual
wavefront distortion. Since AO systems can only partially correct the
wavefront errors, a residual noise component (``speckle noise'') remains
in the corrected PSF. Because of the nature of the AO correction,
this speckle noise does not arise from statistically independent photons.
Instead, speckle-noise constitutes a correlated noise component, which
is the limiting factor for  high-contrast observations (see Racine et al.\
1999, and references therein).
Figure 1 shows an example of the effect of speckle noise.
It was obtained during a lab experiment with a turbulence simulator and the
VLT adaptive optics system NACO. In the top left and right of Figure 1,
two subsequent images of a point source, which was imaged through
the turbulence simulator, and then AO corrected and observed, are displayed. 
In the lower left, the difference between the two images
on the top is shown. Despite the fact that the ``observing conditions''
(r$_{\rm 0}$, $\tau_{\rm 0}$)
were kept constant, strong residuals due to the speckle noise can be observed.

\subsubsection{The concept of differential imaging}

One way to reduce the residuals is the method of dual- (or differential-)
imaging. Here, two images (either in two orthogonal polarization states
-- see Kuhn et al.\ 2001; Potter 2003 --
or in two neighbouring wavelength regions -- see Racine et al.\ 1999; 
Marois et al.\ 2000) have to be obtained simultaneously.
By taking the difference of these two images, the photons from the bright
star (which does not exhibit a high degree of polarization or strong wavelength
dependent spectral features) effectively cancel out, while the much fainter
signal from, e.g., a circumstellar disk (seen in scattered, i.e., polarized
light) or a substellar companion (with strong molecular absorption bands)
clearly stands out. In the lower right panel of Figure 1 the
reduced residuals are apparent (Brandner \& Potter 2002).

\begin{figure}[hbt]
\centerline{
\psfig{figure=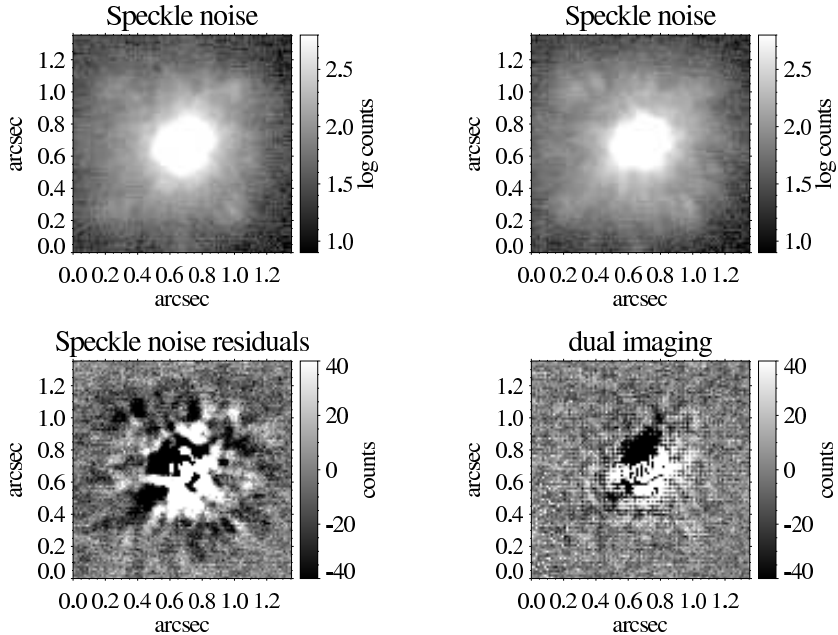,width=11.0cm,angle=0}
}
\caption{Examples of speckle noise, and the gain in noise 
suppression by using dual imaging with a Wollaston prism 
(lab experiment using the VLT AO system NAOS \& CONICA).}
\end{figure}

\begin{figure}[hbt]
\psfig{figure=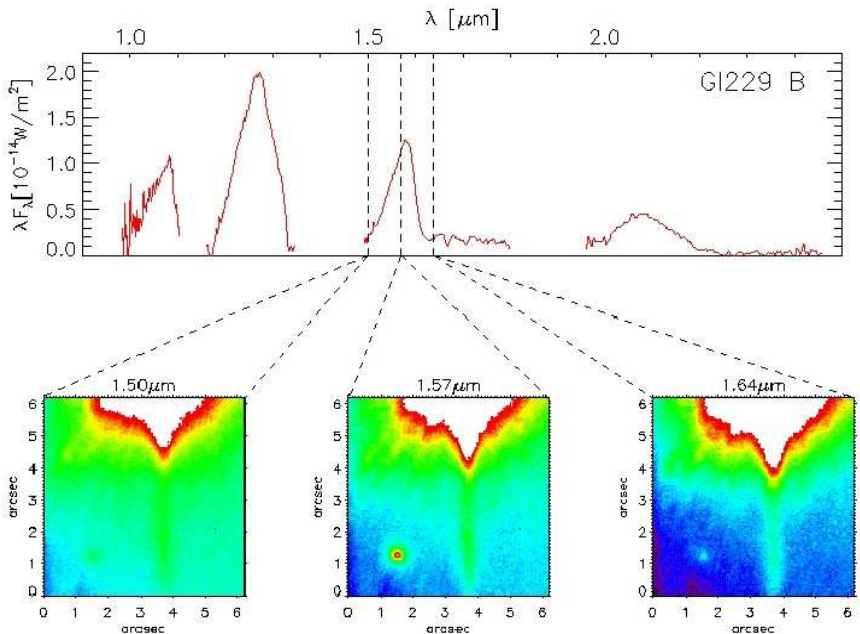,width=12.0cm,angle=0}
\caption{Top: Low-resolution spectrum from the data presented in 
the discovery paper on Gl 229B by Oppenheimer et al.\ (1995). Bottom: Gl 229\,B
observed with the adaptive optics system ADONIS at the ESO 3.6m telescope
with a circular variable filter centered on absorption bands and in the
continuum (Brandner et al.\ 1998). In the ``continuum'' (1.57\,$\mu$m),
Gl\,229\,B is about 2\,mag brighter than at neighbouring wavelengths
($\approx$1.64\,$\mu$m) in a strong CH$_4$ absorption band.}
\end{figure}

As an example, we show in Figure 2 the
brown dwarf Gl\,229b. On the top, the near infrared spectrum obtained
by Oppenheimer et al.\ (1995) is plotted. The shape 
of the spectral energy 
distribution is dominated by strong molecular absorption bands
and the intermittent pseudo-continuum. In the H-band, such an 
intermittent pseudo-continuum is located between water and ammonia 
absorption bands
at wavelengths $\le$ 1.58\,$\mu$m, and a methane absorption band at
wavelengths $\ge$ 1.61\,$\mu$m. In the lower half of Figure 2 
narrow-band imaging data on Gl\,229b as obtained with the
AO system ADONIS at the ESO 3.6m telescope in 1997 is shown. 
A circular variable filter was used  to scan through different wavelength 
ranges. As is apparent from the
images, Gl\,229b is 2\,mag brighter at a wavelength of $\approx$1.57\,$\mu$m
than at the wavelength of neighbouring molecular absorption bands
(see Rosenthal et al.\ 1996).

\subsubsection{Spectral Differential Imagers}

For CONICA, the Spectral Differential Imager (SDI) upgrade 
(led by R.\ Lenzen at MPIA \& L.\ Close at Steward Obs.)
consists of a four-channel beamsplitter combined with a quad-filter.
Two of the quadrants of the quad-filter are adjusted to the longest
wavelength located in a Methane absorption band at 1.625\,$\mu$m, while 
the other two quadrants transmit light at 1.600\,$\mu$m (location of
pseudo-continuum emission) and at 1.575\,$\mu$m (location of a potential 
ammonia absorption band),
respectively. The final differential image is obtained by rescaling,
interpolation and subtraction, similar to the procedure outlined
by Marois et al.\ (2000). 
The implementation of the SDI mode, and the first on-sky tests with
NACO at the VLT were carried out in August 2003, and
a brightness contrast ratio
of $5 \times 10^4$ at a separation of 0.5$''$ from a bright reference source
was achieved.
One of the science goals of the CONICA Planet Finder upgrade is to detect
giant planets with masses down to $\approx$5\,M$_{\rm Jup}$ around nearby 
young stars.

\subsubsection{Atmospheric refraction and spectroscopy}

\begin{table}[htb]
\caption{Differential atmospheric refraction in mas in the JHK bands for median
atmospheric conditions at Paranal (T = 11.5\,C, P = 743\,hPa,
relative humidity 15\%). For comparison, the diffraction limit of
an 8m-telescope is given in the last row. \label{tab1}}
\begin{center}
\begin{tabular}{c|ccc} \hline
wavelength [$\mu$m]  &  1.15--1.35 &        1.55--1.80   &         2.05--2.35 \\ \hline
zenith distance  &  [mas]                 & [mas]              & [mas]   \\ \hline
10$^\circ$&          9        &         4       &            2       \\
20$^\circ$&         18        &         9       &            5       \\
30$^\circ$&         28        &        15       &            8       \\
40$^\circ$&         41        &        21       &           11       \\
50$^\circ$&         59        &        30       &           16       \\
60$^\circ$&         85        &        44       &           23       \\ \hline
FWHM      &         38        &        52     &    68  \\
(1.22$\lambda$/D)    &          &          &      \\
\end{tabular}
\end{center}
\end{table}

Differential atmospheric dispersion leads to elongated (spectrally spread)
PSFs. It scales with the airmass and the overall atmospheric conditions
(see Table 1). For broad-band imaging applications, the apparent
positions of blue and red objects are shifted with respect to each other
along the parallactic angle. This effect, when not corrected for, limits the
accuracy of relative astrometric measurements. Spectroscopy with a narrow slit
(width close to the diffraction limit), which is not aligned with the
parallactic angle, results in a wavelength dependent loss of light. It
is important to keep this in mind when trying to do spectro-photometry.
A much more thorough discussion of the particular effects of long-slit
spectroscopy combined with adaptive optics is presented by Goto (2004).
Many of these effects can also be avoided by employing filled aperture 
integral field spectrographs instead of long-slit spectrographs.

\section{Science Examples}

Science studies of star forming environments can be sub-divided into two
major categories: high-resolution and high-contrast studies of the
immediate environment of a (bright) source, and high-resolution ``wide-field''
studies of a complex or crowded environment. In the following, a brief --
and necessarily incomplete -- summary of star formation studies with AO
is presented.

\subsection{``Wide-field'' AO (\O \ $\le$ 20$''$)}

\begin{figure}[htb]
\psfig{figure=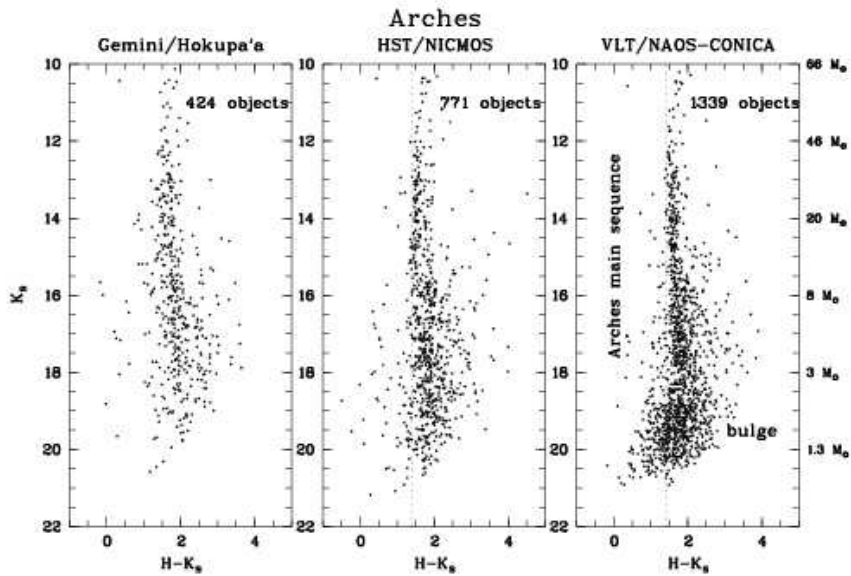,width=12.0cm,angle=0}
\caption{K vs.\ H-K colour-magnitude diagrams for the Arches
cluster as obtained with the low-order curvature AO system Hokupa'a at
Gemini North, NICMOS aboard HST, and the intermediate order
Shack-Hartmann AO system
NACO at the ESO VLT (Stolte et al.\ 2003).}
\end{figure}

Studies of stellar populations in crowded fields require high-angular
resolution. One area of particular interest is the mass function of
Galactic and extra-galactic starburst clusters. Examples of starburst
clusters studied with AO include R136 in the 30 Doradus region in the Large
Magellanic Cloud (Brandl et al.\ 1996), the central cluster of the
giant HII region NGC 3603 in the Carina spiral arm (Eisenhauer et al.\ 1998),
and the Arches cluster in the Galactic Center region (Blum et al.\ 2001, 
Stolte et al.\ 2002, Yang et al.\ 2002).

Figure 3, taken from Stolte et al.\ (2003), shows a comparison of 
ground- and space based NIR observations
of the Arches cluster. In the case of low Strehl ratios (left), crowding 
limits the sensitivity towards faint point sources. The colour-magnitude
diagram on the right is based on data with moderate SR of 14\% in H, and 20\% in
K, and clearly demonstrate that ground-based AO at 8\,m-class telescopes
can be superior to HST/NICMOS (middle panel).
The Orion star forming region has also been studied with AO,
in particular the stellar population in the central core of the Trapezium 
cluster (Simon et al.\ 1999), photoevaporating
stellar envelopes (McCullough et al.\ 1995), and the
complex interstellar environment in the vicinity of the BN/KL object
(e.g., Vannier et al.\ 2001).

\subsection{High-contrast AO (Contrast $\ge$ 10$^5$)}

Despite the above mentioned limitations due to PSF variations and
speckle noise, AO has been very successfully employed to study 
the close environment of young stellar objects. This includes both 
a search for faint point
sources like substellar companions, and for faint extended emission from
circumstellar disks, envelopes and outflows.
Stellar companions to young stars like NX Pup (Brandner et al.\ 1995),
VW Cha (Brandeker et al.\ 2001), HD\,98800 (Prato et al.\ 2001),
$\chi$\,1\,Orionis (K\"onig et al.\ 2002) or
TY CrA (Chauvin et al.\ 2003) have been identified by
means of AO.
Liu et al.\ (2002), e.g., detected a substellar (spectral type L)
companion to the young solar analog HR\,7672 (see Figure 4),
and a binary  brown dwarf companion to the young star HD\,130948
was identified by Potter et al.\ (2002) and Goto et al.\ (2002).
Systematic studies of the binary content of star forming regions like
IC 348 (Duch\^ene et al.\ 1999), the Pleiades (Mart\'{\i}n et al.\ 2000),
NGC 6611 (Duch\^ene et al.\ 2001), MBM 12 (Chauvin et al.\ 2002),
NGC 2024 (Beck et al.\ 2003) have been carried out as well.

\begin{figure}[hbt]
\centerline{
\hbox{
\psfig{figure=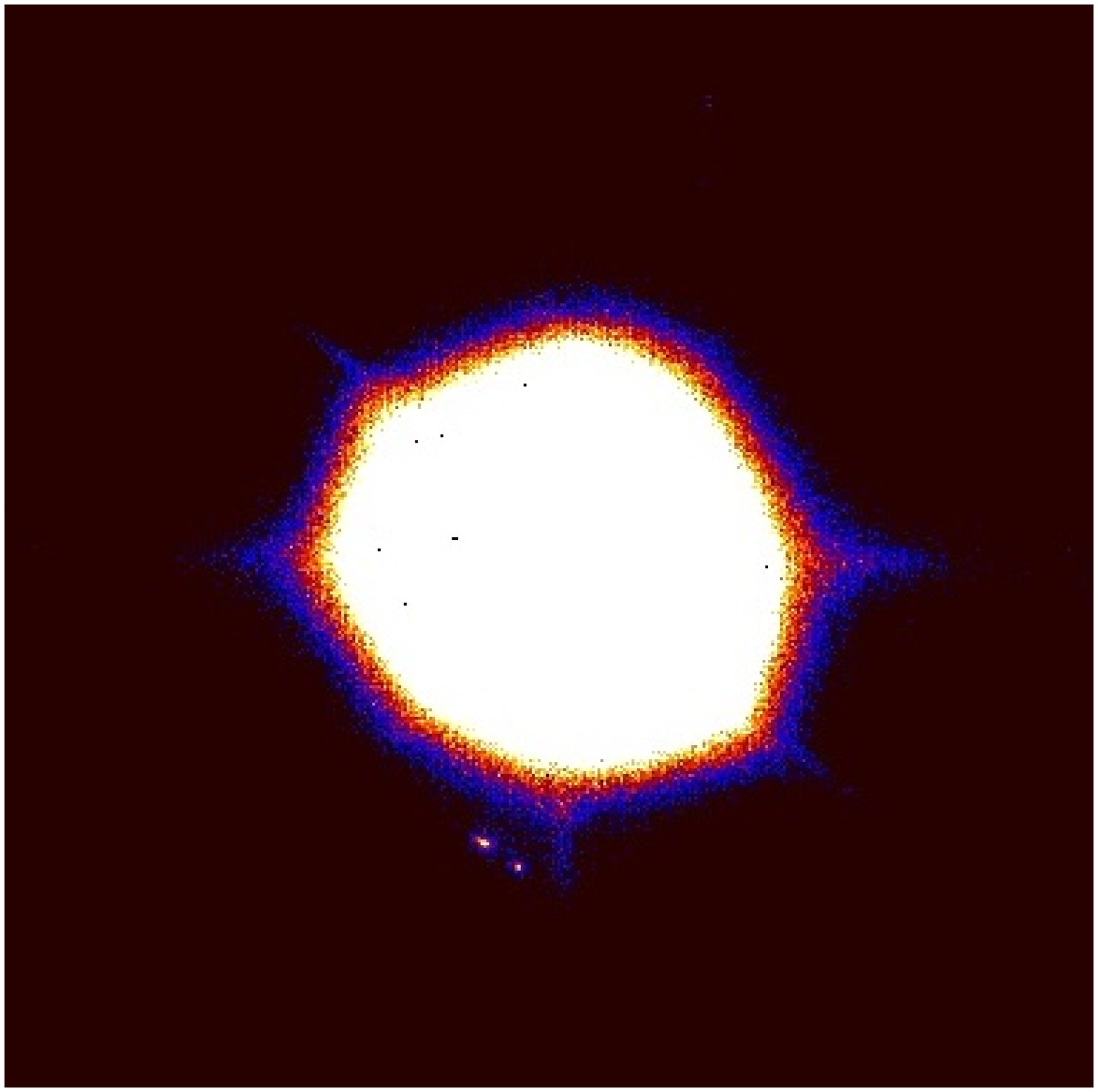,width=5.0cm,angle=0}
\psfig{figure=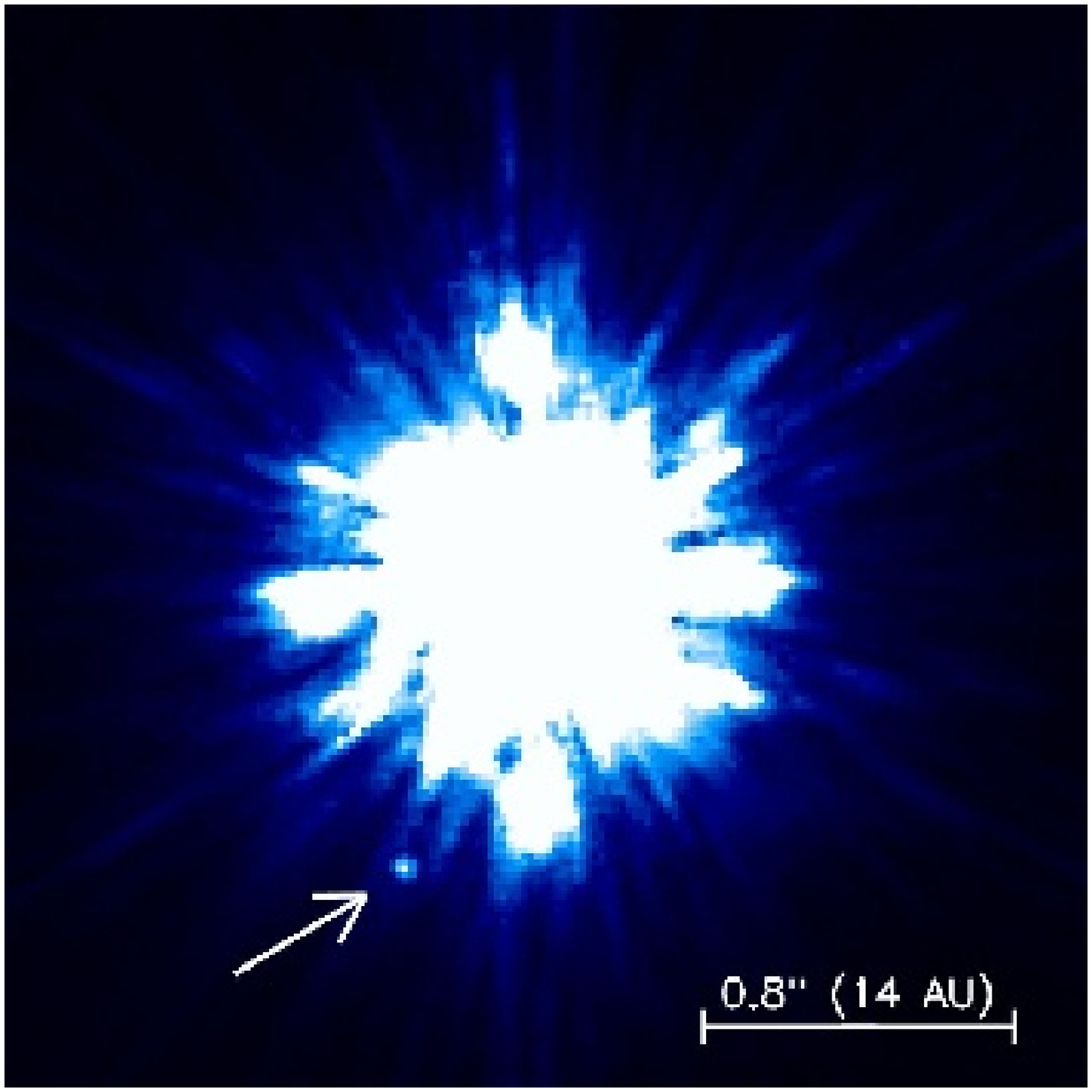,width=5.0cm,angle=0}
}}
\caption{Faint companions and companion candidates:
The Palomar AO system (left) detected this binary companion candidate,
which has a separation of 5$''$, and is about 13\,mag fainter in K than the 
primary (courtesy of Matthew Britton, Caltech). Using the Keck AO system (right),
Liu et al.\ (2002) detected an L-dwarf companion to the young solar
analog HR 7672. The companion is at
a separation of 0.79$''$, and 8.6\,mag fainter than the primary in K-band.}
\end{figure}

Jets and outflows from young stars were studied by Dougados et al.\ (2000), 
Millan-Gabet \& Monnier (2002), and L\'opez-Martin et al.\ (2003).
Ultra-compact H\,II regions are studied, e.g., by Feldt et al.\ (1999)
and Henning et al.\ (2001).
Quite a number of AO studies also concentrated on the properties of 
circumstellar disks around young stars such as GG\,Tau (Roddier et al.\ 1996,
Itoh et al.\ 2002), HD\,100546 (Pantin et al.\ 2000),
UY Aur (Potter et al.\ 2000),  LkH$\alpha$\,198 (Fukagawa et al.\ 2002),
HD 141569 (Boccaletti et al.\ 2003), HD 150193\,A (Fukagawa et al.\ 2003),
HV Tauri\,C (Stapelfeldt et al.\ 2003),
or led to the detection of edge-on disks, like, e.g., in MBM 12 (Jayawardhana
et al.\ 2002), or around HV\,Tau\,C (Monin \& Bouvier 2000).

\subsection{Spectroscopy with AO}

T\,Tauri itself has been the focus of near-infrared spectroscopic studies 
combined with adaptive optics in the past couple of years. Kasper et al.\
(2002) obtained H- and K-band spectra of the main components T\,Tauri North
and South using the Calar Alto 3.5\,m telescope with its AO system
ALFA, while Duch\^ene et al.\ (2002) took K-band spectra of the two 
components comprising the
close binary T\,Tauri South A\&B using the Keck AO system.
Using the PUEO system at CFHT, Garcia et al.\ (1999) obtained spatially
resolved spectroscopy of the Z\,CMa components. Davies et al.\ (2001)
studied young stellar objects in LkH$\alpha$\,225 by means of AO integral
field spectroscopy with the ALFA system on Calar Alto.

\subsection{Interferometry and AO}

Both the Keck and the VLT Interferometer are now equipped with AO systems.
The first science results obtained on the circumstellar disk around DG\,Tau
have just been published (Colavita et al.\ 2003, see also Akeson, these
proceedings), and many more results are expected for the near future.

\section{Further Reading}

For further reading, the following
online references on the theory and application
of adaptive optics are recommended:
\begin{itemize}
\item[$\bullet$] AO Lecture by Laird Close: \\
http://athene.as.arizona.edu/$\sim$lclose/a519/Lecture\_1.html
\item[$\bullet$] AO Lecture by Claire Max: http://cfao.ucolick.org/$\sim$max/289C
\item[$\bullet$] (Theory of) AO tutorial by Andrei Tokovinin:\\
http://www.ctio.noao.edu/$\sim$atokovin/tutorial/intro.html
\item[$\bullet$] AO: Past, Present and Future by Olivier Lai: \\
http://www.cfht.hawaii.edu/Instruments/Imaging/AOB/local\_tutorial.html
\end{itemize}


\begin{references}

\reference Beck, T.L., Simon, M., Close, L.M.\ 2003, ApJ 583, 358
\reference Blum, R.D., Schaerer, D., Pasquali, A.\ et al.\ 2001, AJ 122, 1875
\reference Boccaletti, A., Augereau, J.C., Marchis, F., Hahn, J.\ 2003, ApJ 585, 494
\reference Brandeker, A., Liseau, R., Artymowicz, P., Jayawardhana, R.\ 2001,
ApJ 561, L199
\reference Brandl, B., Sams, B.J., Bertoldi, F.\ et al., 1996, ApJ 466, 254
\reference Brandner, W., Bouvier, J., Grebel, E., Tessier, E., De Winter, D.,
Beuzit, J.\ 1995, A\&A 298, 818
\reference Brandner, W., Frink, S., K\"ohler, R., Kunkel, M.\ 1998, in {\it
10th Cambridge Workshop on Cool Stars, Stellar Systems and the Sun}, 
eds.\ R.A.\ Donahue \& J.A.\ Bookbinder, ASP Conf.\ Ser.\ 154, p.\ 1836
\reference Brandner, W., Potter, D.\ 2002, in {\it Scientific Drivers for ESO
Future VLT/VLTI Instrumentation}, eds.\ J.\ Bergeron \& G.\ Monnet (Berlin:
Springer-Verlag), 264
\reference Chauvin, G., M\'enard, F., Fusco, T.\ et al.\ 2002, A\&A 394, 949
\reference Chauvin, G., Lagrange, A.-M., Beust, H.\ et al.\ 2003, A\&A 406, L51
\reference Colavita, M., Akeson, R., Wizinowich, P.\ et al.\ 2003, ApJ 592, L83
\reference Davies, R.I., Tecza, M., Looney, L.W.\ et al.\ 2001, ApJ 552, 692
\reference Dougados, C., Cabrit,S., Lavalley, C., M\'enard, F., 2000, A\&A 357,
L61
\reference Duch\^ene, G., Bouvier, J., Simon, T.\ 1999, A\&A 343, 831
\reference Duch\^ene, G., Simon, T., Eisl\"offel, J., Bouvier, J.\ 2001, A\&A 379, 147
\reference Duch\^ene, G., Ghez, A.M., McCabe, C.\ 2002, ApJ 568, 771
\reference Eisenhauer, F., Quirrenbach, A., Zinnecker, H., Genzel, R.\ 1998,
ApJ 498, 278
\reference Feldt, M., Stecklum, B., Henning, T., Launhardt, R., Hayward, T.L.\ 
1999, A\&A 346, 243
\reference Fukagawa, M., Tamura, M., Suto, H.\ et al.\ 2002, PASJ 54, 969
\reference Fukagawa, M., Tamura, M., Itoh, H.\ et al.\ 2003, ApJ 590, L49
\reference Garcia, P.J.V., Thiebaut, E., Bacon, R.\ 1999, A\&A 346, 892
\reference Goto, M., Kobayashi, N., Teradam H.\ et al.\ 2002, ApJ 567, L59
\reference Goto, M.\ 2004, in {\it ESO workshop on 
``Science with Adaptive Optics''}, eds.\ W.\ Brandner \& M.\ Kasper (Berlin:
Springer-Verlag), in press
\reference Henning, T., Feldt, M., Stecklum, B., Klein, R.\ 2001, A\&A 370, 100
\reference Jayawardhana, R., Luhman, K.L., D'Alessio, P., Stauffer, J.R.\ 2002, ApJ 571, L51
\reference Kasper, M.E., Feldt, M., Herbst, T.M., Hippler, S., Ott, T., Tacconi-Garman, L.E.\ 2002, ApJ 568, 267
\reference K\"onig, B., Fuhrmann, K., Neuh\"auser, R., Charbonneau, D., Jayawardhana, R.\ 2002, A\&A 394, L43
\reference Kuhn, J.R., Potter, D., Parise, B.\ 2001, ApJ 553, L189
\reference Liu, M.C., Fischer, D.A., Graham, J.R., Lloyd, J.P., Marcy, G.W., Butler, R.P.\ 2002, ApJ 571, 519
\reference L\'opez-Martin, L., Cabrit, S., Dougados, C.\ 2003, A\&A 405, L1
\reference Marois, C., Doyon, R., Racine, R., Nadeau, D.\ 2000, PASP 112, 91
\reference Mart\'{\i}n, E.L., Brandner, W., Bouvier, J.\ et al.\ 2000, ApJ 543, 299
\reference McCullough, P.R., Fugate, R.Q., Christou, J.C.\ et al.\ 1995, ApJ
438, 394
\reference Millan-Gabet, R., Monnier, J.D.\ 2002, ApJ 580, L167
\reference Monin, J.-L., Bouvier, J.\ 2000, A\&A 356, L75
\reference Oppenheimer, B.R., Kulkarni, S.R., Matthews, K., Nakajima, T.\ 1995, Science 270, 1478
\reference Pantin, E., Waelkens, C., Lagage, P.O.\ 2000, A\&A 361, L9
\reference Potter, D.E., Close, L.M., Roddier, F.\ et al.\ 2000, ApJ 540, 422
\reference Potter, D., Mart\'{\i}n, E.L., Cushing, M.C., Baudoz, P., Brandner,
W., Guyon, O., Neuh\"auser, R.\ 2002, ApJ 567, L133
\reference Prato, L., Ghez, A.M., Pi\~na, R.K.\ et al.\ 2001, ApJ 549, 590
\reference Racine, R., Walker, G.A.H., Nadeau, D., Doyon, R., Marois, C.\ 1999, PASP 111, 587
\reference Roddier, F., Roddier, C., Northcott, M.J., Graves, J.E., Jim, K.\
1996, ApJ 463, 326
\reference Rosenthal, E.D., Gurwell, M.A., Ho, P.T.P.\ 1996, Nature 384, 243
\reference Simon, M., Close, L., Beck, T.L.\ 1999, AJ 117, 1375
\reference Stapelfeldt, K.R., M\'enard, F., Watson, A.M.\ et al.\ 2003, ApJ 589,
410
\reference Stolte, A., Grebel, E., Brandner, W., Figer, D.\ 2002, A\&A 394,
459
\reference Stolte, A., Brandner, W., Grebel, E., Figer, D.F.\ et al.\ 2003,
The Messenger No.\ 111, p.\ 9
\reference Tessier, E., Bouvier, J., Beuzit, J.-L., Brandner, W.\ 1994,
The Messenger No.\ 78, p.\ 35
\reference Vannier, L., Lemaire, J.L., Field, D., Pineau des Forets, G.,
Pijpers, F.P., Rouan, D.\ 2001, A\&A 366, 651
\reference Yang, Y., Park, H.S., Lee, M.G., Lee, S.-G.\ 2002, JKAS 35, 131

\end{references}
\end{document}